# Residual Nanoscale Strain in Cesium Lead Bromide Perovskite Reduces Stability and Alters Local Band Gap

[1]Materials Science & Engineering Program, University of California, San Diego, La Jolla, CA 92093

[2]Department of Nanoengineering, University of California, San Diego, La Jolla, CA 92093

[3]Advanced Photon Source, Argonne National Laboratory, Argonne, IL 60439

Xueying Li[1], Yanqi Luo[2], Martin Holt[2], Zhonghou Cai[2] and David P. Fenning[1,2]

**Corresponding Author**

*David P. Fenning. Email: dfenning@eng.ucsd.edu

ABSTRACT. Using nanoprobe X-ray diffraction microscopy, we investigate the relationship between residual strains from crystal growth in $CsPbBr_3$ thin film crystals, their stability, and local bandgap. We find that out-of-plane compressive strain that arises from cooldown from crystallization is detrimental to material stability under X-ray irradiation. We also find that the optical photoluminescence red shifts as a result of the out-of-plane compressive strain. The sensitivity of bandgap to strain suggests possible applications such as stress-sensitive sensors. Mosaicity, the formation of small misorientations in neighboring crystalline domains we observe in some $CsPbBr_3$ single crystals, indicates the significant variations in crystal quality that can occur even in single-crystal halide perovskites. The nano-diffraction results suggest that reducing local strains is a necessary path to enhance the stability of perovskite optoelectronic materials and devices from light-emitting diodes to high-energy detectors.

## INTRODUCTION

The manipulation and control of strain during crystal growth is foundational for advanced optoelectronic materials and heterostructures. The effect of strain on band gap and mobility has been intensely studied and exploited in Si, Ge, and III-Vs,[1-3] while strain engineering is a key enabler of complex oxide perovskite heterostructures.[4,5] In contrast, understanding of strain and control of it during growth of the emergent halide perovskites is still limited. With long carrier diffusion lengths and a tunable band gap as a function of composition,[6,7] the halide perovskites have shown promise for diverse optoelectronic applications including solar cells,[8] light emitting dioides,[9] lasing,[10] and photodetectors.[11] Despite their facile processing and significant strides in optimization of optoelectronic properties, halide perovskites still suffer from poor durability with respect to traditional semiconductors – perhaps the major barrier to their commercialization.

Strain is one of the causes of low-crystallinity in halide perovskite thin film[12] and nanoparticles.[13] The commonly-used spin casting and annealing method is a fast crystallization process, which impacts the carrier lifetime and mobility of the perovskite layer.[14] Controlling the fabrication process to enhance the crystallinity has been found to effectively improve the perovskite-based device performance. A number of studies have recently shown that the improved crystallinity with proper substrates,[15,16] halide intermediates[17] and metal doping[18] gives better perovskite solar cell performance. In the recent study of $CH_3NH_3PbI_3$ thin films by Zhao *et al.*, strain created by the thermal mismatch with the substrate during crystallization annealing after spin-casting was found to be detrimental to device stability.[12] Avoiding thermal annealing by using solvent exchange to crystallize the perovskite has been shown to reduce strain and improve perovskite stability.[19] These recent studies shed new light on the important role of strain and stability in halide perovskite. To better resolve the impact of strain in halide perovskite materials and devices, localization of strains in perovskite thin films is necessary to precisely attribute the effect of strain on stability.

The development of halide perovskites as high-energy photodetectors is also gaining attraction due to their high detectivity.[20–23] Specification of perovskite stability under diverse functional environments including high-energy irradiation and identification of the factors that limit stability is needed for deployment of halide perovskite across optoelectronic applications.

Beyond stability, strain may play a significant role in the optoelectronic properties of halide perovskites. First-principles density functional theory studies have proposed that hydrostatic pressure or biaxial strain can reduce the band gap in halide perovskites.[24–26] Several experimental investigations have studied the effect of hydrostatic pressure on band gap in hybrid ($APbX_3$ with A=$CH_3NH_3^+$ (MA) or $CH(NH_2)_2^+$ (FA) and X=Br or I)[27–33] and inorganic cesium lead halide perovskite, $CsPbBr_3$.[34] The band gap narrows under <1 GPa hydrostatic compression due to increasing in antibonding overlap of Pb-X; on the contrary, the band gap broadens under high pressure (>1 GPa), which could be due to phase change or amorphization in the high-pressure regime.[29] Hall *et al.* also observed a blue shift of photoluminescence (PL) of $CH_3NH_3PbI_3$ in the presence of humidity, and their first-principles computation shows a simultaneous increase of lattice constant and band gap when intercalating water into the lattice,[35] suggesting that the band gap may be tunable by straining the crystal lattice. Along similar lines, researchers have recently demonstrated an increase in band gap with epitaxial growth of $CsPbBr_3$ on $VO_2$ nanowires, which results from compressive *in-plane* strains in the $CsPbBr_3$ during expansion of the nanowires.[36] While these studies introduce strains in extreme environments or characterize the average bulk strain in perovskite films, detection of the strain distribution from common fabrication processes and understanding of the nanoscale effects of strain is needed to provide detailed insight for feedback on synthesis strategies and for designing new mechanically-responsive functionality in these optoelectronic materials.

Here we use nanoprobe X-ray Diffraction Microscopy (nanoXRD) to quantify the nanoscale strain distribution in $CsPbBr_3$ thin film single crystals and relate these local strains to a detrimental effect on stability and a locally-varying bandgap. Leveraging a strain gradient introduced during cooldown from crystal growth within a given crystal provides the opportunity to study nuanced effects of strain on stability and bandgap within the same single crystal, avoiding sample-to-sample variations including minor variations in growth or subsequent environmental conditions. We find that nanoscopic regions with higher out-of-plane compressive strains exhibit decreased stability and a redshift in bandgap and that pronounced mosaicity can arise during growth of perovskite thin film single crystals.

**Experimental**

**Synthesis of $CsPbBr_3$ crystals:** $CsPbBr_3$ crystals are grown following a previously-reported method that forms thin film single crystals tens of microns in size[37] while imitating the common spin casting and annealing method. The inorganic precursors CsBr (powder, 99% Sigma Aldrich) and $PbBr_2$ (powder, 98% Sigma Aldrich) are used as purchased. 0.33 M 1:1 $CsBr/PbBr_2$ was dissolved in dimethylsulfoxide. The substrates are sonicated in soap water, IPA and water. The substrates are treated by oxygen plasma before deposition. Using patterned Pt strips on quartz to create a coefficient of thermal expansion (CTE) mismatch, a lateral strain gradient is formed in CsPbBr$_3$ crystals that grow from solution spanning the Pt and quartz substrate. For the Pt-patterned quartz, 100-nm thick Pt is sputtered onto the quartz. The perovskite precursor solution is spin-cast onto the substrate at 500 rpm for 10 seconds. The substrate is immediately pressed onto the PDMS stamp facing down and annealed at 100 °C for 2 minutes, after which they are cooled on the fume hood bench. The CsPbBr$_3$ crystals under investigation fall across pre-patterned Pt strips on glass, as shown schematically in Figure 1a.

**Nanoprobe X-ray Diffraction:** Scanning X-ray nano-diffraction microscopy (nanoXRD) and X-ray fluorescence (XRF) were performed using the Hard X-ray Nanoprobe (HXN) of the Center for Nanoscale Materials (CNM) at sector 26-ID-C of the Advanced Photon Source, Argonne National Laboratory. NanoXRD uses a focused beam with 220 Bragg diffraction geometry and rasters the X-ray beam across the crystal with the scattering recorded on a pixel-array detector (Medipix) at every point to produce a map of nanoscale strain. A Fresnel zone plate (Xradia Inc., 133-μm diameter gold pattern, 24-nm outer zone, 300-nm thickness) focuses the monochromatic incident X-ray beam (photon energy 9.0 keV, λ=1.378 Å) onto the sample with a ≈50-nm full-width half-maximum lateral beam cross-section in the focal plane with a photon flux of 3x10$^8$ photons/sec.[38] The perovskite sample is mapped with a 1.5-$\mu$m stepsize while the pixel array detector captures the diffracted intensity at each scan position. A single-element florescence detector is used to locate the crystal over the Pt electrode edge on quartz. To image the scattered intensity from the (220) peak of the orthorhombic CsPbBr$_3$ crystal, the sample is positioned at 220 Bragg diffraction to the X-ray incident beam, and the detector is positioned at 2θ=27.44° (λ=1.3777 Å). A map of 2θ variation is obtained by calculating the centroid of 2θ in each diffraction image. The scattering intensity and position in 2θ is extracted from each image. The strain in the [220] direction is calculated from the *d*-spacing, *d'*, extracted from the measured 2θ diffraction at each spot relative to the literature (220) d-spacing, *d°*, of the room-temperature orthorhombic CsPbBr$_3$ phase (ICSD 97851) as in Eq. 1:

$$\varepsilon_{220} = \frac{d'_{220} - d^o_{220}}{d^o_{220}} \qquad (1)$$

The instrument maintains angles between the X-ray beam, sample, and detector with ± 0.001° precision. In addition, the diffraction detector spans a width of 1° 2θ with a pixel resolution of 0.003°. This state-of-the-art control provided by the instrument allows us to observe the strain distribution precisely and accurately at the nanoscale. When lattice plane rotation is present, we eliminated the rotation error and extracted the magnitude strain by simulating the diffraction pattern with Patterson function,[39] which separates the shift of the bright diffraction fringe from the movement of the aperture projection in the diffraction image. Since there is minimal rotation (no shift of aperture center) detected in the strained crystal, the 2θ is converted directly to d-spacing and then strain relative to the literature value.

NanoXRD also provides an opportunity to study the stability and degradation mechanism of perovskite crystals under intense ionizing irradiation, relevant to potential applications in high-energy detection. The $6\times10^{17}$ photons/s-cm$^2$ synchrotron X-ray irradiation generates energetic core holes and hot electrons that, after a cascading thermalization process, settle at the band edges of the perovskite. The X-ray dose is calculated from the flux $3\times10^8$ photons/sec and beam diameter 240 nm at the 220 Bragg angle to the sample surface. Each hard X-ray photon absorbed in the perovskite generates >10$^3$ band edge carriers, according to the empirical relationship $N_{carriers} = E_{X-ray}/3E_g$.[40] After accounting for partial X-ray transmission through the nanometric perovskite film, the nano-focused X-ray irradiation is equivalent to ~700 suns, providing an accelerated test condition to monitor the stability and degradation of the perovskite crystal in the presence of a large concentration of carriers.[41]

To analyze any structural degradation, a correlation coefficient of the diffraction pattern at each frame with respect to the initial one was calculated as a representation of structural degradation of perovskite, according to:

$$r = \frac{\sum_m \sum_n (A_{mn} - \bar{A})(B_{mn} - \bar{B})}{\sqrt{(\sum_m \sum_n (A_{mn} - \bar{A})^2)(\sum_m \sum_n (B_{mn} - \bar{B})^2)}} \qquad (2)$$

in which $A_{mn}$ is the signal count of one pixel in the 514-by-514 initial diffraction image, and $B_{mn}$ is the signal count of the same pixel of the subsequent diffraction image. $\bar{A}$ and $\bar{B}$ are the average signal counts of the initial diffraction image and the subsequent one respectively. The time when the $r$ drops below 0.9 was select for comparison ($t_{90}$). To assess the effect of strain while preserving the sample properties, each spot for time series analysis was separated from the last by at least 1 $\mu$m. With a beam FWHM about 50 nm, the 1-$\mu$m spacing is large enough to minimize irradiation-induced damage carryover to nearby scan spots.

**Micro-photoluminescence:** Photoluminescence mapping was performed at 50x magnification using a Renishaw inVia microscope equipped with a Si CCD array detector. A 514-nm laser source with 1.0 $\mu$W power was scanned in 1 $\mu$m steps. Luminescence was collected via a 1200 mm$^{-1}$ grating, which with the 514-nm incident laser results in a spectral resolution about 0.1 nm. To better match the same sampling volumes between the normal-incidence PL maps and the nano-diffraction that samples the whole thickness of the 2 $\mu$m thick crystal at an oblique angle, an 8-μm horizontal averaging filter was applied to PL maps. Additional experimental details can be found in the supplemental Information.

**Results and Discussion**

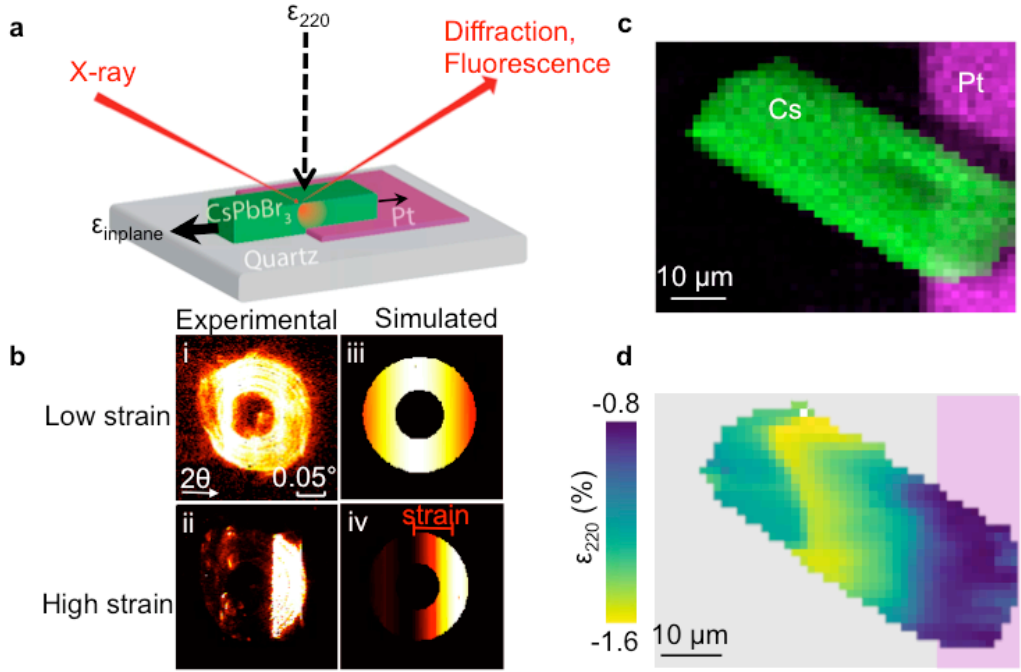

**Figure 1.** (a) Schematic of the nano-diffraction measurement of a strain gradient induced in a perovskite thin film single crystal by the CTE mismatch at the quartz and Pt substrate interfaces beneath it. The nano-XRD quantifies the out-of-plane strain component only, $\varepsilon_{220}$. (b) Diffraction images of experimental diffraction patterns of (i) low-strain and (ii) high-strain regions in crystal. Simulated diffraction patterns of (iii) low-strain and (iv) high-strain regions in crystal. The shift of the bright vertical fringe on the circular aperture is caused by strain. (c) Plan view X-ray fluorescence map of Cs (green) and Pt (purple) counts, indicating the crystal's position across the quartz/Pt edge. (d) Experimental out-of-plane strain map in a $CsPbBr_3$ crystal characterized by nanoXRD $CsPbBr_3$ on quartz/Pt substrate.

Local changes in *d*-spacing due to strain are obtained from each diffraction image by the relative position of the diffraction fringe on the circular aperture. Figure 1b shows a diffraction image from a location on the $CsPbBr_3$ exhibiting a low-strain position of the bright fringe (Figure 1b(i)) compared to a diffraction image where the position of the bright diffraction fringe on the circular-shaped aperture has moved to higher 2θ, indicating a more compressively-strained lattice (Figure 1b(ii)). To confirm and quantify this shift due to strain, Patterson function simulation of the diffraction pattern is used (Figure 1b(iii, iv)). The low-strain and high-strain diffraction patterns in Figure 1b(i and ii) are calculated to exhibit -1.02% and -1.59% strain respectively. This strain is separated from the movement of the whole circular aperture on the diffraction detector, which can result from the rotation of the lattice planes within the crystal (e.g. if the surface is curved). With minimal tilt of the crystal as evidenced by the stationary aperture projection in the diffraction images and the flatness of the crystal as measured by AFM (Figure S1), the 2θ variation is almost exclusively the result of strain in the crystal.

The X-ray fluorescence (XRF) map of the $CsPbBr_3$ crystal spanning the Pt electrode and quartz substrate is shown in Figure 1c, where the Cs corresponds to the perovskite crystal and the Pt signal is from the electrode. If we assume that the growing $CsPbBr_3$ does not freely slip from the substrate interface, the contraction of $CsPbBr_3$ during cooling from crystallization is restricted by the relatively lower CTEs of the Pt and glass substrate. The $CsPbBr_3$ crystal is thus expected to exhibit residual tensile strain in-plane and a resultant compressive strain out-of-plane, as governed by its Poison's ratio (ν=0.33).[42] Since the crystal is deposited across two distinct materials (Pt and quartz) with lower CTEs than the perovskite, the CTE difference between Pt and quartz should result in a strain gradient across the Pt edge as shown schematically in Figure 1a. Assuming *no* slip at the substrate-crystal interface, a finite element simulation of the 2-μm thick crystal shows the crystal is compressively strained overall as indicated by a negative total integrated out-of-plane displacement everywhere along the length of the crystal as shown in Figure S2a (see Table S1 for CTEs used in simulation). As a result of the difference in CTEs, the crystal is expected to be more strained on the quartz than Pt (Figure S2a). A simulated cross-section of the out-of-plane component of the strain distribution can be found in Figure S2b.

Nano-diffraction microscopy reveals that the entire crystal has a compressive out-of-plane strain, and the region above the exposed quartz substrate is more compressively strained relative to the area above the Pt, as expected from simulation. The nano-diffraction map of strain in the $CsPbBr_3$ crystal is shown in Figure 1d, captured simultaneously with the X-ray fluorescence (XRF) map shown in Figure 1c. The spatially-resolved, out-of-plane strain measured via nano-X-ray diffraction varies between -0.8% and -1.6%, indicating that the crystal is compressively strained in the out-of-plane direction. In general agreement with the trends from simulation, the magnitude of the strain decays toward the left and right-hand edges of the crystal and is highest nearer the quartz/Pt interface. The average lateral strain gradient across the crystal generated by the two different substrates is about 0.03%/µm. In the center of the crystal, the strain varies smoothly from higher to lower compressive strain from left to right due to the CTE mismatch of the perovskite with the Pt and quartz.

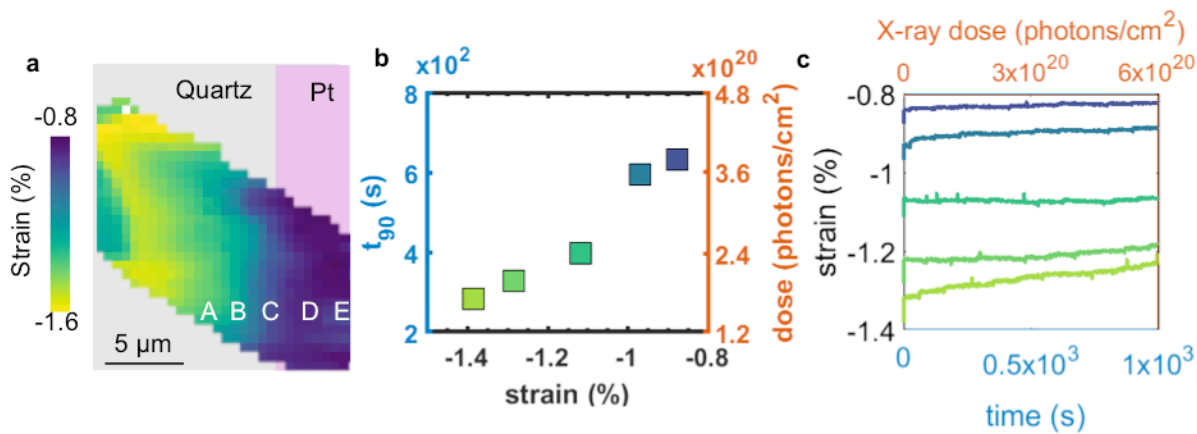

**Figure 2.** (a) Five measurement locations along the strain gradient designated A- E were selected for time series analysis of the diffraction pattern. (b) The time over which the two-dimensional correlation coefficient of the diffraction pattern decays to 90% of its initial value ($t_{90}$) is plotted versus the local strain at locations A-E. (c) At each location A-E, the compressive strain is shown as a function of time and X-ray dos, indicating the relaxation of strain during structural degradation under X-ray irradiation.

To study the effect of strain on stability under X-ray irradiation, time series measurements were taken on different spots along the strain gradient on the crystal shown in Figure 2a and labeled positions A-E. In each time series measurement, the spot was exposed to the high-flux X-ray beam continuously to observe the change in the X-ray scattering over time. The two-dimensional correlation coefficient of the diffraction image relative to the initial image is used as an indicator of degradation. The obtained correlation coefficient as a function of time under x-ray can be found in Figure S3. Movies that visualize the sequence of diffraction pattern images at each spot are available in the Supplemental Information.

We find that positions with less residual strain exhibit increased structural robustness under X-ray irradiation, as indicated by a longer decay time of the scattered intensity. Figure 2b plots the relationship between strain state and $t_{90}$, the time when the correlation coefficient of the diffraction pattern drops below 0.9, that is when scattering image has changed more than 10%. Less-strained spots have a longer $t_{90}$. The degradation decay lifetime appears to vary only due to the different strain magnitudes, as spots D and E are atop Pt and A-C are atop the quartz but they fall roughly colinearly in Figure 2b. This direct correlation between structural decay time and strain indicates that nanoscale residual strains negatively affect the stability of halide perovskites.

During the degradation process under X-ray irradiation, the compressive strain is relaxed over time, as shown in Figure 2c. This observation indicates that the generation of structural defects under intense X-ray irradiation simultaneously lowers the diffraction pattern correlation coefficient and relaxes residual strain, consistent with previous observation of strain relaxation during degradation in $CH_3NH_3PbI_3$.[12] Nano-XRD time series studies of hybrid perovskite $CH_3NH_3PbBr_3$ crystals reveal that the hybrid perovskites undergo much more rapid structural degradation compared to $CsPbBr_3$ (Figure S4), as expected given their relatively poorer thermal and environmental stability and low enthalpy of formation.[43] The $t_{90}$ of $CH_3NH_3PbBr_3$ corresponds to an X-ray dose of $10^{18}$ photon/cm$^2$ at 9 keV at an estimated flux of $3 \times 10^8$ photons/s, while the $t_{90}$ of $CsPbBr_3$ corresponds to an X-ray dose of $10^{20}$ photon/cm$^2$. These dose limits provide bounds for the design of perovskite-based high energy particle detectors.

The process by which the perovskite crystal degrades is evident from the time-series diffraction patterns (Figure S4b and Supplemental Movies). The time series from $CH_3NH_3PbBr_3$ (MAPbBr$_3$) crystals reveal that the crystal structure collapses on itself locally as degradation proceeds. Upon extending the X-ray dose beyond >10x that required to collect a single high-quality scattering image, the MAPbBr$_3$ (002) $d$-spacing decreases and the angular spread in the $\chi$-direction normal to $2\theta$ increases. The spreading of the reflection in the $\chi$-direction indicates that the crystal planes are bending. Together, the shrinking of the out-of-plane spacing and simultaneous warping of the (001) planes indicates a collapse of the crystal planes inward. This collapse proceeds until scattering fades completely and the local structure of the perovskite is amorphized.

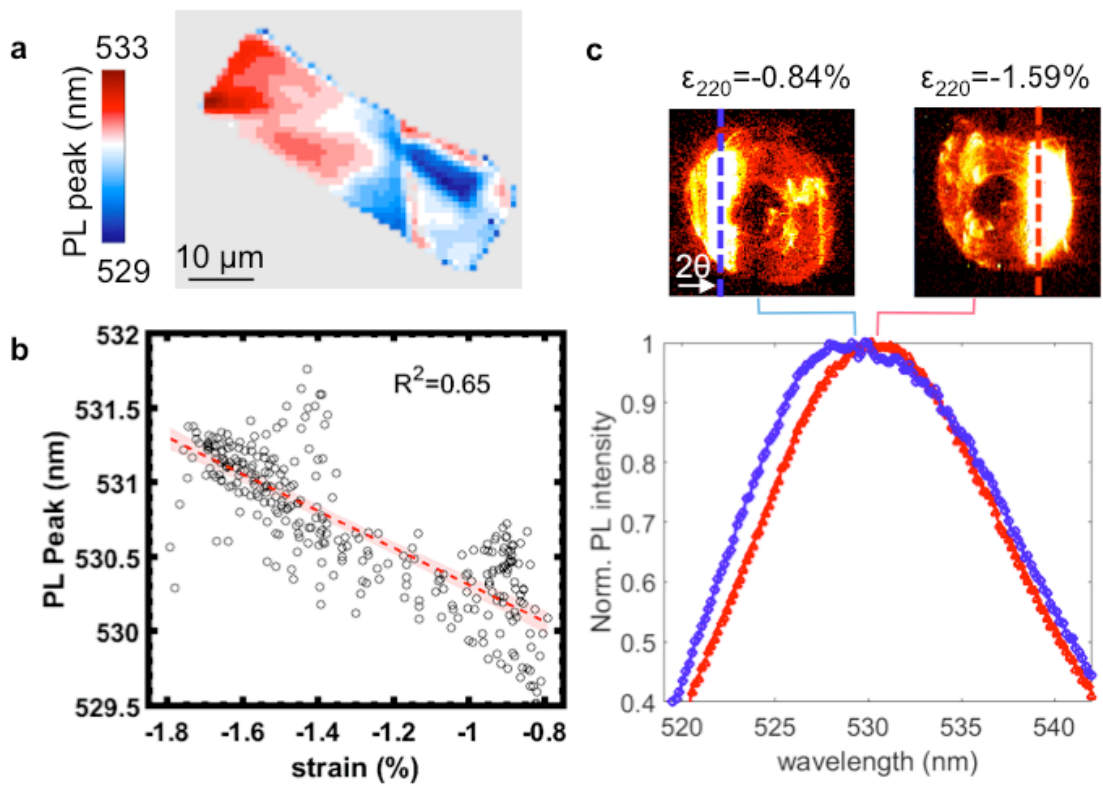

**Figure 3.** (a) Map of the wavelength of peak PL intensity for the same $CsPbBr_3$ crystal measured by nano-XRD and shown in Figure 1-2. (b) Negative correlation of the local PL peak versus nanoscale strain. The red line indicates the best fit line, and the shaded area is the functional prediction interval on the linear regression. (c) Examples of diffraction images (top) at spots on the crystal with strain -0.84% (blue) and -1.59% (red) and corresponding PL spectra (bottom). The diffraction peak shifts to higher $2\theta$ at the more compressed location as denoted by the dashed line. The spot with larger compressive out-of-plane strain has a red-shifted PL peak.

Micro-photoluminescence ($\mu$-PL) mapping of the $CsPbBr_3$ crystal measured by NanoXRD reveals a negative correlation between local strain and the peak wavelength of PL emission. The PL peak generally varies from longer to shorter wavelengths moving top left to bottom right along the crystal (Figure 3a), save a clear discontinuity at the Pt/glass interface and some edge effects. By comparison with the strain map from nano-XRD (Figure 1d), this blue shift of PL left to right across the crystal corresponds with the trend of moving from a region of higher compressive strain on the top left of crystal to lower compressive strain at the bottom right. We make this relationship explicit in the scatter plot in Figure 3b, where a clear negative correlation is seen between the point-by-point strain and PL peak wavelength in this crystal ($R^2=0.65$). A single Gaussian was used to extract the PL peak at each point as shown in Figure S5. Figure 3c highlights two spots that exemplify this trend with local strains of -0.84% and -1.59% determined from NanoXRD, where a red shift at higher compressive strain is seen in their local PL emission. Extrapolating from the dataset, the unstrained crystal is expected to have a PL peak at 529 nm, in agreement with previously reported values.[44] The PL peak position shifts 1.2 +/-0.1 nm per 1% strain, suggesting that perovskites may provide sensitive opto-mechanical sensing.

The red shift concomitant with compressive out-of-plane strain may be due to at least two physical processes. One factor is that the strained $PbBr_6$ octahedra change the Pb-Br anti-bonding overlap and thus results in the red shift.[29] Another factor could be structural defects generated at strained regions leading to the band gap narrowing.[45] These shifts in bandgap are not the result of the appearance of secondary phases. The powder XRD only contains peaks from orthorhombic phase of $CsPbBr_3$ (Figure S6a), and each nanoscopic step of the beam across the crystal sample diffracts at the 2θ expected from the [220] peak of o-$CsPbBr_3$. The nano-XRF map shows the Cs:Pb ratio is uniform across the whole crystal area, which excludes the possibility of Cs:Pb stoichiometric variation (Figure S6b). Furthermore, electron backscatter diffraction (EBSD) identifies the crystal as orthorhombic $CsPbBr_3$ everywhere (Figure S7). Thus, the observed PL red shift can only be ascribed to the measured variation in local strain.

However, the correlation between PL peak and strain is weakened here because of several experimental factors. First, the PL and the nano-diffraction are most sensitive to different sampling volumes. The PL is most sensitive to the top 230 nm, the $1/e$ attenuation depth of the incident 514 nm laser in $CsPbBr_3$,[46] whereas the full thickness is probed in X-ray diffraction. Second, the substrate-induced strain is largest in magnitude near the substrate interfaces and relaxes through the crystal thickness in the vertical direction even in the absence of structural defects, as indicated by the simplified mechanical model simulated in Figure S2b. Third, the crystal stamping method used here to produce the thin film single crystal does not yield the structural quality of bulk single crystals. Thus, we expect the fundamental underlying relationship between band gap and strain to be stronger than that reported here.

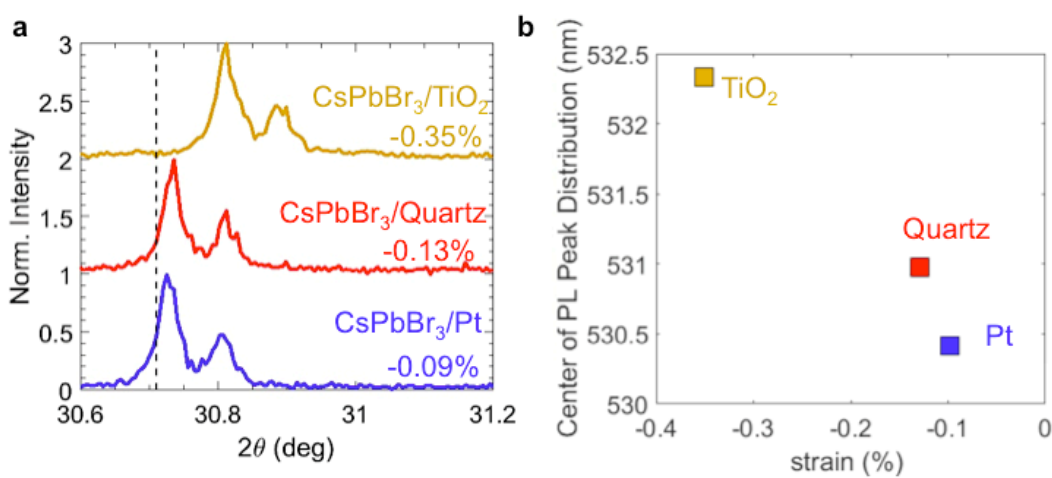

**Figure 4.** (a) Powder XRD patterns of the (220) peak of CsPbBr$_3$ crystals on Pt, quartz and TiO$_2$ substrates with the corresponding strain values. The dashed line is the literature powder XRD value of 220 peak (ICSD 97851) (b) Center of mass of the PL peak distribution in each crystal versus strain measured by benchtop XRD on Pt, quartz and TiO$_2$ substrates.

Based on the clear link between PL peak and residual strain from our nanoscale characterization, we were able to reproduce the trend in polycrystalline perovskite thin film preparations on various substrates including Pt, quartz and compact TiO$_2$ (Figure 4). With high-angular resolution benchtop XRD and careful sample height alignment, we determined the lattice constant in the [220] out-of-plane direction of CsPbBr$_3$ crystals on each substrate and calculate the strain according to Eq. 1. The CsPbBr$_3$ crystals have the largest compressive strain in the out-of-plane direction on TiO$_2$ followed by quartz and are least strained on Pt. The full XRD patterns (Figure S8) show the crystals are orthorhombic Pbnm phase on all three substrates. The corresponding peak wavelength of PL emission decreases as compressive strain decreases from TiO$_2$ to Quartz to Pt to with values of 532.3, 531.0, and 530.4 nm respectively. The higher compressive strain on quartz than Pt is consistent with the trend of their CTE mismatch with the perovskite, but the commonly-used compact TiO$_2$ substrate shows higher compressive strain despite having a CTE near Pt. It is possible that the wetting interaction between the perovskite and the nanoparticulate TiO$_2$ reduces the slip or strain relaxation at the interface between the perovskite crystals and the substrate, resulting in higher strain during crystallization. Despite benchtop thin-film XRD averaging strain many crystallites due to its large spot size, a small negative trend of PL peak versus strain is still evident, although the benefit of nanoscale characterization of model materials is clear in elucidating this relationship between strain and the peak wavelength of photoluminescence emission.

Perhaps surprisingly, the local nano-XRD scattering intensity does not correlate with local PL intensity, contrary to convention in the physical chemistry of semiconductors that higher crystallinity and lower crystallographic defect density corresponds to enhanced photoluminescence (Figure S9). This observation may be partially explained by the sensitivity of the PL to the unpassivated top surface while the nano-XRD probes through the thickness. However, it may also be an indication the self-passivating nature of defective perovskite regions.[47,48] Recent work has observed that amorphous regions at grain boundaries in MAPbBr$_3$ thin films may be passivating and display increased PL lifetime and intensity.[49] Further investigation of the relationship between perovskite crystallinity on the nanoscale and their opto(mechano)electronic properties is warranted.

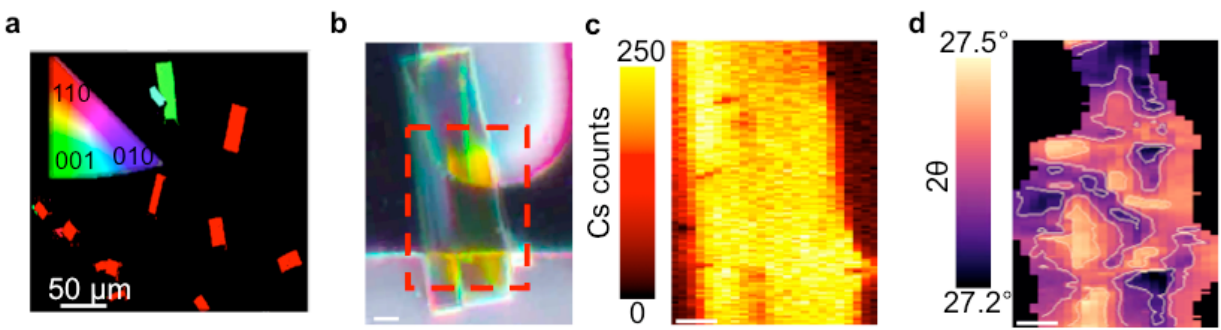

**Figure 5.** (a) As-fabricated CsPbBr$_3$ crystals appear as single-crystals with Pbnm (110) orientation in electron backscatter diffraction measurements. (b) Optical image of a CsPbBr$_3$ single crystal, where the red box indicates the region of nano-XRD mapping; (c) Plan view X-ray fluorescence map of Cs counts, showing the mapped crystal area. (d)The less than 0.3° variation in 2θ measured by nanoXRD reveals mosaicity. Scale Bar is 10 $\mu$m in (b), (c) and (d).

Finally, we note that not all of the stamped CsPbBr$_3$ perovskite crystals exhibit a single crystalline domain upon nanoprobe investigation. Although they appear as uniform, single crystals under the more common crystallographic analysis of electron-back scatter diffraction (EBSD) (Figure 5a), the crystal in the optical image of Figure 5b showed significant mosaicity when investigated by nanoXRD. In Figure 5c, the Cs XRF map shows the extent of the crystal, but only some areas diffract due to mosaicity, as characterized by a map of the angle of peak diffraction intensity producing by rocking curve measurement (Figure 5d). The mosaics are diffracting crystalline domains within the larger crystal with slight misorientation. When the misorientation angle is small (<0.3°), the domain diffracts and displays a small shift in 2θ due to the crystal plane rotation (Figure 5d). As noted above, this rotation is straightforwardly separated from strain in analysis of the diffraction image. In contrast, where the misorientation is large, the diffraction is not detected by the diffraction detector and appears dark in the scattering intensity map (Figure 5d). In this sample, mosaicity dominates the nanoscale structure of the film, rending detailed strain analysis impossible due to the distorted and twinned diffraction patterns (Figure S10). The formation of crystal mosaics may be due to the low formation energy of perovskite crystals and the rapid crystallization process.[50,51] Scrutinizing the crystallinity at the nanoscale, we find that mosaicity can easily result during perovskite growth. Mosaicity is also expected to reduce the stability and increase the variation in optoelectronic properties of perovskite films due to the reduction in structural quality. To our knowledge, this is the first report of mosaicity in halide perovskite crystals, though its detrimental impacts on stability have been previously hypothesized.[52] Further understanding of the heterogeneous distribution of strain including mosaicity is necessary to understand how to improve the crystallinity of perovskite materials and requires local nanoscale characterization of crystal structure.

**Conclusion**

In summary, we study the effect of nanoscale residual strain on halide perovskite crystal structural stability and band gap using synchrotron nanoprobe X-ray diffraction. Using a strain gradient that develops within $CsPbBr_3$ thin film single crystals due to the thermal expansion mismatch between the perovskite crystal and substrates, we find reduced durability at locations of higher compressive out-of-plane strain. By using thin film single crystals, we remove any effect of grain boundaries, demonstrating the intrinsically detrimental effect of such strain on the stability of halide perovskites. Time-series analysis of structural degradation under X-ray irradiation indicates an imploding collapse of the crystal structure and provides quantitative dose limits for perovskite application in high-energy detectors. With micro-photoluminescence mapping, we correlate the local optoelectronic properties with the measured nanoscale strain. From -0.8 to -1.6% strain, the PL peak is continuously red-shifted from 529 nm to 533 nm. The fine and continuous tuning of bandgap with strain suggests new potential strain sensing applications for perovskites.

Because we use substrates, solution processing, and growth temperatures ubiquitous in the perovskite literature, we emphasize that re-engineering crystal growth to alleviate strain offers the opportunity to stabilize perovskite materials and devices relative to their current standard. Based on the observation that stability decreases monotonically with nanoscale strain, any localized stress concentrations that form during crystallization or in the application of perovskite films in curved or flexible devices may serve as "weakest links" for incipient degradation. Our observation of mosaicity in halide perovskite crystals at the nanoscale also indicates that crystal quality in even single crystals may affect reports of the bulk properties of halide perovskites. Further work is necessary to improve the crystallinity of perovskites at the nanoscale, reduce residual strains, and leverage the unique optoelectromechanical properties of the perovskites. Deeper understanding of how the strain is distributed at the nanoscale in perovskite materials will allow exploitation of the relationship between structural deformation and shifts in optoelectronically-active energy levels.

ASSOCIATED CONTENT

**Supporting Information**. Additional experimental details and tabulated coefficients of thermal expansion. Simulated strain distribution in a $CsPbBr_3$ crystal on Pt/quartz substrate. Time series of the diffraction intensity decay of $CsPbBr_3$ under X-ray irradiation. Example of fitting PL spectrum. XRF, powder XRD, EBSD, and AFM of $CsPbBr_3$ crystal. PL intensity map and XRD intensity map. Example nano-diffraction patterns of areas with complex microstructure. Movies of $CsPbBr_3$ and $MAPbBr_3$ diffraction pattern as a function of time under X-ray irradiation. The following files are available free of charge.

SupplementalInformation.pdf

AUTHOR INFORMATION

Corresponding Author

*Email: dfenning@eng.ucsd.edu

**Notes**

The authors declare no competing financial interests.

ACKNOWLEDGMENT

The authors thank Joseph M. Palomba for setting up the height alignment in powder XRD measurement, and Tyler Harrington and Kevin Kaufmann for EBSD measurement. This research was supported by a Hellman Fellowship and the California Energy Commission EPC-16-050. Use of the Center for Nanoscale Materials and the Advanced Photon Source, both Office of Science user facilities, was supported by the U.S. Department of Energy, Office of Science, Office of Basic Energy Sciences, under Contract No. DE-AC02-06CH11357.

**Table of Content Artwork**

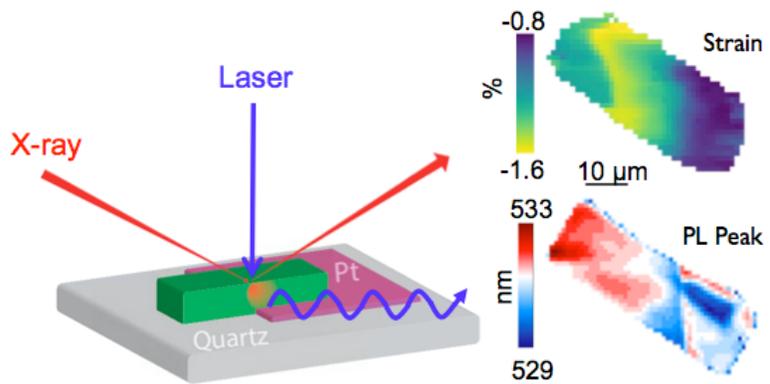

Residual Nanoscale Strain in Cesium Lead Bromide Perovskite Reduces Stability and Alters Local Band Gap

Xueying Li[1], Yanqi Luo[2], Martin Holt[2], Zhonghou Cai[2] and David P. Fenning[1,2]

[1]Materials Science & Engineering Program, University of California, San Diego, La Jolla, CA 92093

[2]Department of Nanoengineering, University of California, San Diego, La Jolla, CA 92093

[3]Advanced Photon Source, Argonne National Laboratory, Argonne, IL 60439

Supporting Information

**Electron Backscatter Diffraction**
To identify the crystal phase, electron backscatter diffraction (EBSD) is measured on the crystal using an FEI Apero scanning electron microscopy and Oxford symmetry detector at 20 kV. The crystal structures of orthorhombic $CsPbBr_3$ ICSD # 97851, tetragonal $CsPbBr_3$ ICSD # 109295 and cubic $CsPbBr_3$ ICSD # 29073 were used to match the crystal phase.

**Atomic Force Microscopy**
The AFM image was obtained using a Veeco Scanning Probe Microscopy in tapping mode and scanning over a range of 50 μm by 70 μm at a resolution of 360 × 512 data points.

**Benchtop X-ray diffraction**
Powder XRD spectra were taken by a Bruker AXS D8 Discover diffractometer in parallel beam geometry with Cu Kα radiation (λ=1.5418 Å). The height alignment accessory was used for accurate strain calculation.

Supporting Information

| Materials | CTE ($10^{-6}$ K$^{-1}$) |
|---|---|
| CsPbBr$_3$[1] | 37.7 |
| Pt[2] | 9.0 |
| Quartz[3] | 0.55 |
| TiO$_2$[4] | 2.57 |

Table S1. Coefficients of thermal expansion of CsPbBr$_3$, Pt, quartz and TiO$_2$

Supporting Information

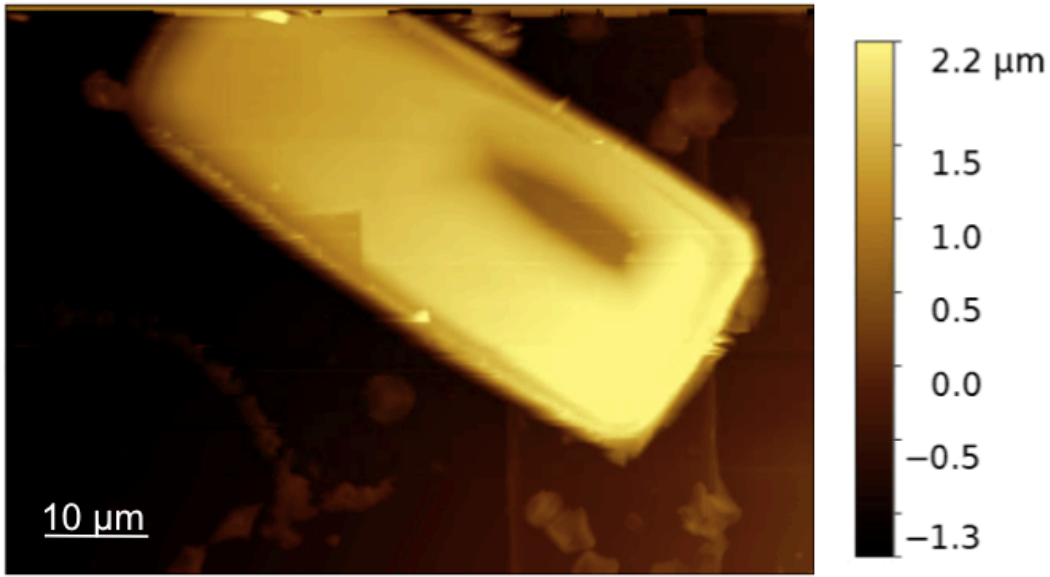

Figure S1. AFM of the CsPbBr$_3$ crystal measured by nanoXRD shows the crystal is 2-um thick with a flat smooth surface except for one large void on the right of the crystal. The roughness Ra is around 40 nm on the crystal excluding the void.

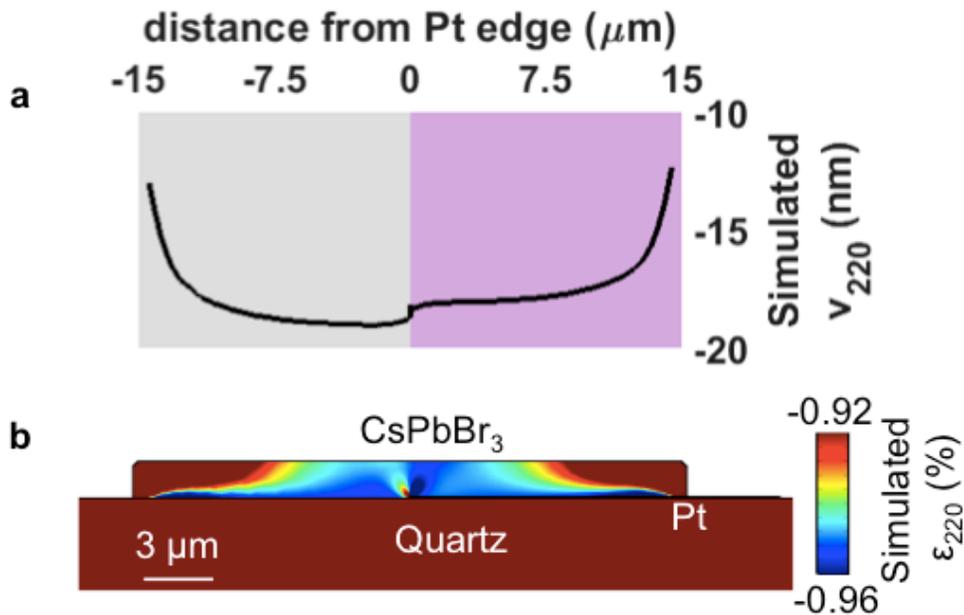

Figure S2. (a) Simulated displacement of a 2-um thick CsPbBr$_3$ crystal on Pt-patterned quartz substrate. (b) COMSOL mechanical simulation of the out-of-plane strain in a the CsPbBr$_3$ crystal on Quartz/Pt substrate.

Supporting Information

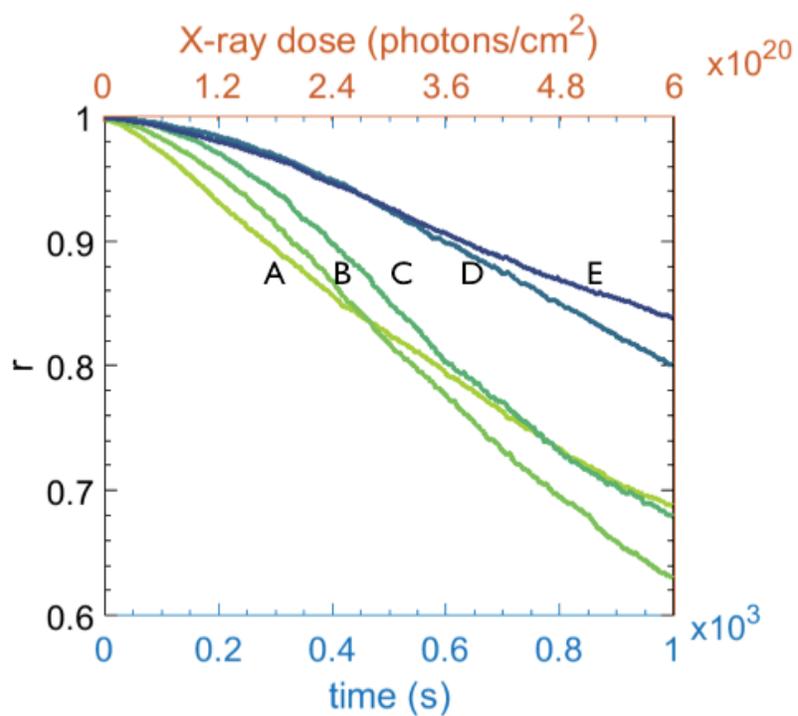

Figure S3. Correlation Coefficient of diffraction images as a function of time under X-ray at five positions on the strained crystal in figure 2.

Supporting Information

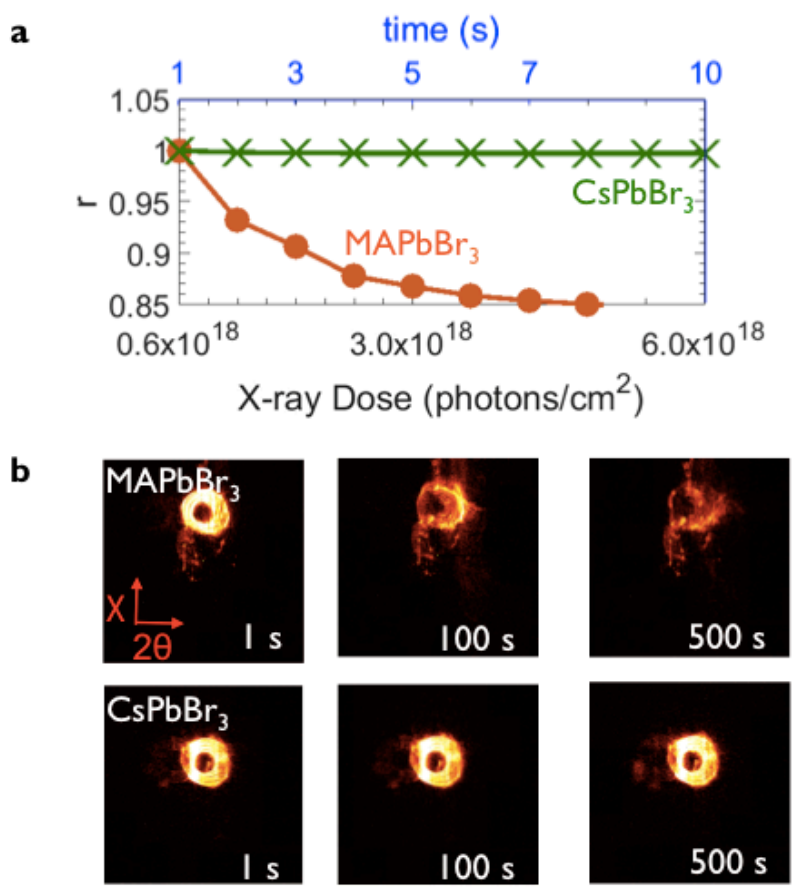

Figure S4. (a) Correlation coefficient r of the diffraction image vs. accumulated X-ray dose shows that CsPbBr$_3$ (green curve) are much more stable than MAPbBr$_3$ (red curve). (b) Diffraction images of MAPbBr$_3$ and CsPbBr$_3$ diffraction at 1, 100 500s showing the relatively stable CsPbBr$_3$ under X-ray irradiation and the loss of long range order in the hybrid perovskite.

Supporting Information

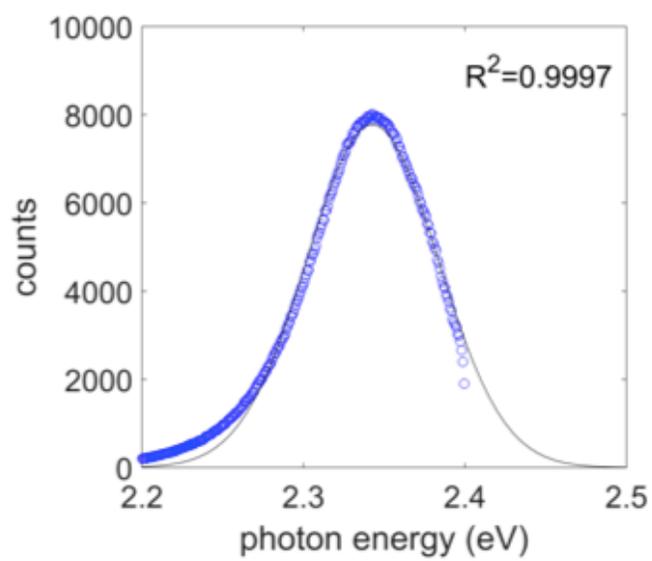

Figure S5. Example of single Gaussian fit of PL spectrum of CsPbBr$_3$ from µ-PL.

Supporting Information

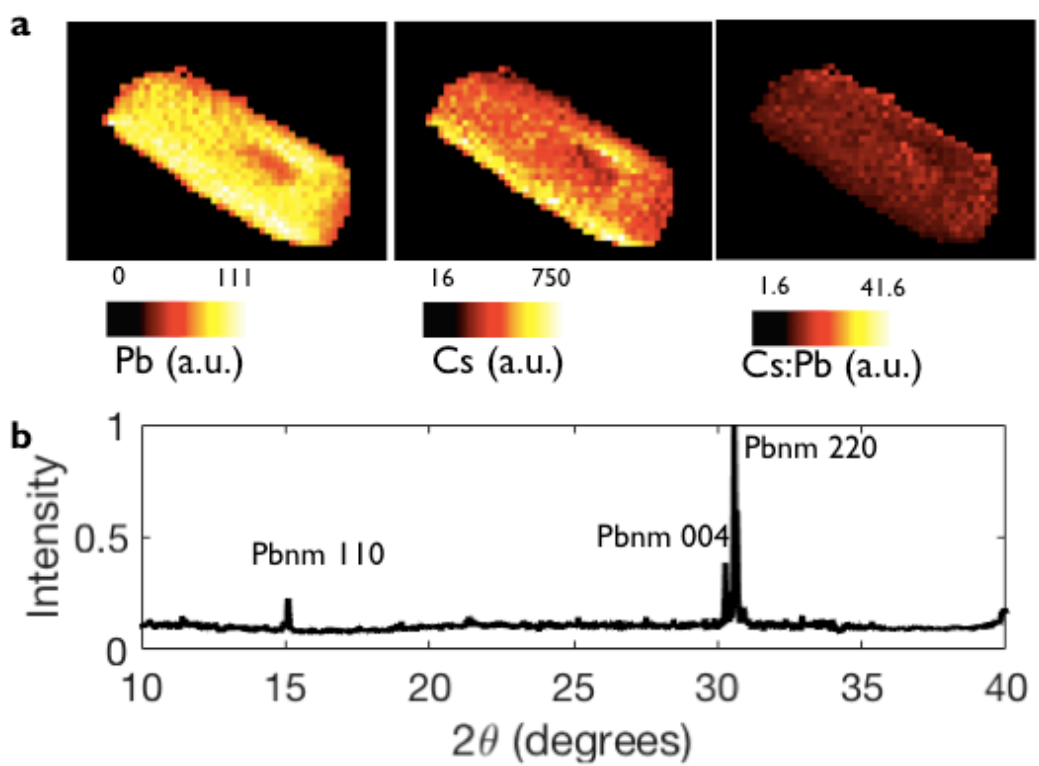

Figure S6. (a) Pb, Cs, Cs/Pb ratio maps in XRF. (b) Powder XRD of the CsPbBr$_3$ sample measured by nanoXRD.

Supporting Information

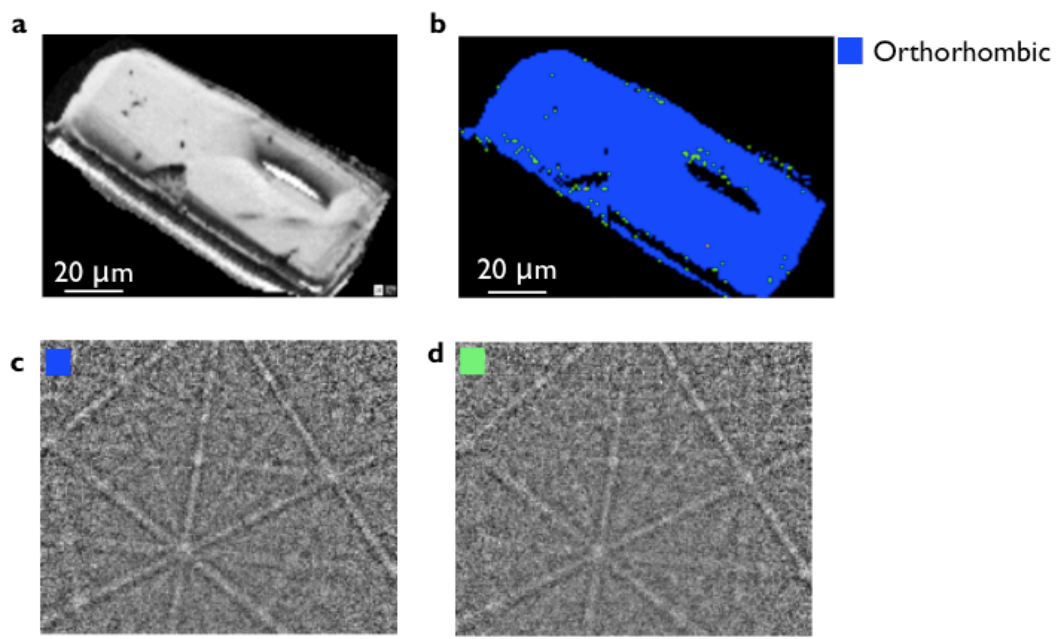

Figure S7. (a) SEM image of the CsPbBr$_3$ crystal measured after nanoXRD. (b) EBSD phase map of the CsPbBr$_3$ crystal shows the crystal is orthorhomic CsPbBr$_3$ with space group Pbnm. The green dots are not automatically identified as orthorhombic phase in the software, due to poorly indexed orthorhombic patterns, as shown by (c) the Kikuchi pattern of a blue point identified as orthorhombic CsPbBr$_3$ and (d) the Kikuchi pattern of a poorly-indexed (green) point identified mistakenly as cubic CsPbBr$_3$. The symmetry and pattern in (d) precisely matches that of (c).

Supporting Information

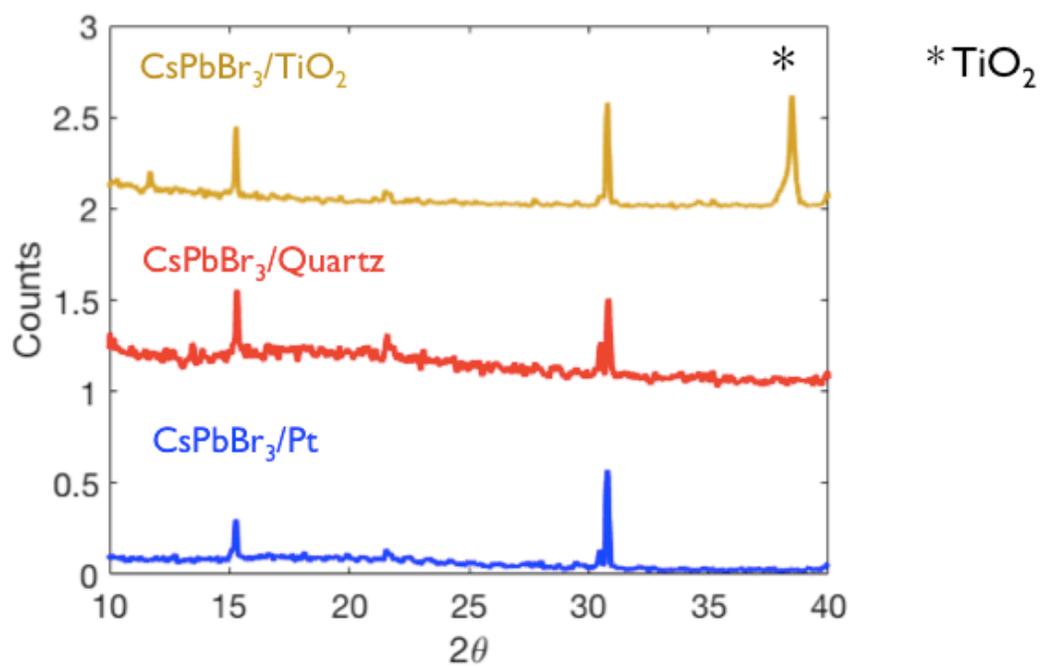

Figure S8. Powder XRD patterns of CsPbBr$_3$ crystals on TiO$_2$, Quartz, and Pt.

Supporting Information

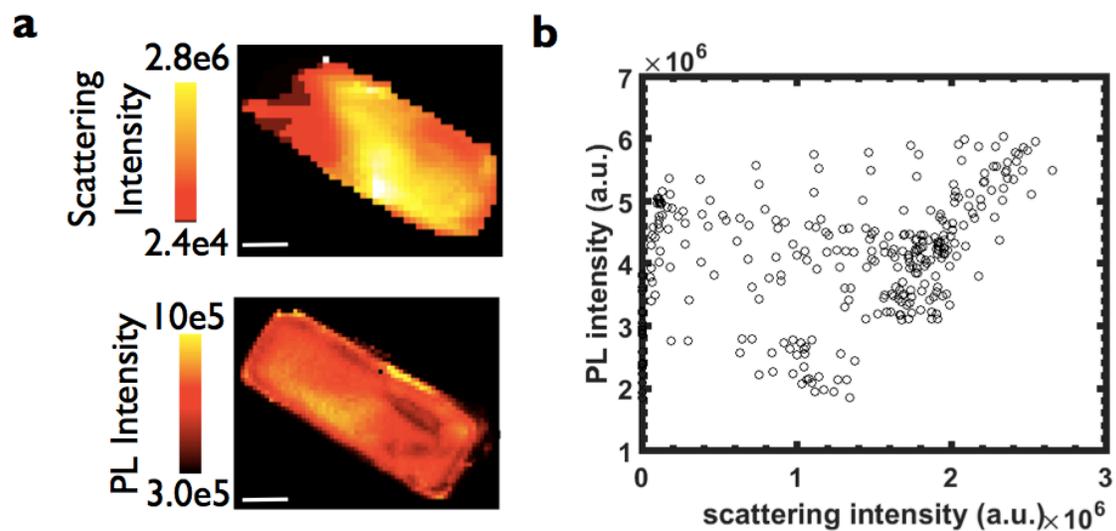

Figure S9. (a) Scattering intensity map from nanoXRD and PL intensity map from microPL on the same $CsPbBr_3$ crystal. (b) Scatter plot of PL intensity versus scattering intensity shows non-association. Scale bar is 10 μm in (a).

Supporting Information

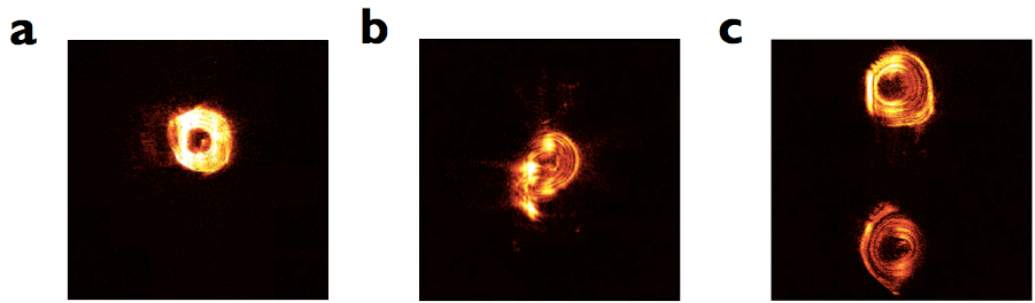

Figure S10. Diffraction images of (a) the homogeneously strained $CsPbBr_3$ crystal from Figure 1, and (b) distortion and (c) twining in the mosaic crystal of Figure 5.

Supporting Information